# THE *IDEAL GAS BEHAVIOR* IS IN FACT NOTHING, BUT A *MACROSCOPIC QUANTUM MECHANICAL MANIFESTATION*

Tolga Yarman, [*] Department of Engineering, Okan University, Istanbul, Turkey & Savronik, Organize Sanayii Bölgesi, Eskisehir, Turkey, tolgayarman@gmail.com

Alexander L. Kholmetskii, Department of Physics, Belarus State University, Minsk, Belarus, khol123@yahoo.com

Jean-Louis Tane, Former Faculty Member, Department of Geology, University Joseph Fourier, Grenoble, France (tanejl@aol.com)

**ABSTRACT**

This article mainly consists in the quantum mechanical study of an adiabatically compressed particle, in an infinitely high well, which we conjecture, can be considered as the basis of an *ideal gas.* Thus we prove that, all the compression energy is, as may be expected, transformed into *extra kinetic energy* of the particle. This result, frames a quantum mechanical definition of an *ideal gas.* It further helps the elucidation of a paradox.

## 1. INTRODUCION

In Section 1 we consider a non-relativistic quantum mechanical particle in an infinitely high potential well. We thus check out that the work done to squeeze adiabatically the size of the box is transformed into extra kinetic energy of the particle in consideration. In Section 2 we handle the same problem, though along with a relativistic quantum mechanical particle. Thus, the work done to squeeze adiabatically the size of the box is transformed into extra relativistic energy. In Section 3, we summarize the *parallelism,* the first author et al. have previously drawn between the classical laws of thermodynamics, mainly the *law of adiabatic expansion,* and *quantum mechanics.* We generalize our findings in Section 4 to come to the conclusion that an *ideal gas* is made of *individual quantum mechanical constituents, imprisoned in the container of the gas,* well yielding to the known macroscopic thermodynamic manifestations. This result is nothing but strikingly, a *macroscopic aspect of quantum mechanics.* Our findings allow us to answer in Section 5, the following essential question, seemingly paradoxical, at a first strike: *If work is achieved between the two compartments of a closed system, would the overall "internal energy", change?* The answer *(given that, there is no heat exchange with the outside, seems to a first strike, be "Yes", but),* is still *"No" (that is, the overall internal energy of the closed system, remains the same).* This is in fact, nothing, but the *law of conservation of energy,* yet checked via a quantum mechanical analysis of an ideal gas, which we were able to define accordingly. Finally a conclusion is drawn in Section 6.

## 1. THE NON-RELATIVISITIC PARTICLE IN THE BOX: THE WORK DONE TO SQUEEZE THE SIZE OF THE BOX, IS FULLY TRANSFORMED INTO ETRA KINETIC ENERGY

We consider a particle of rest mass $m_0$ in an infinitely high box. For simplicity we visualize only a one dimensional box of size L, of $1\,\text{cm}^2$ of lateral surface. Thus the particle is *quantum mechanically* confined in this box, where it oscillates indefinitely, back and forth, along the x axis. The total energy $E_n$ of the particle, first supposed moving at non-relativistic velocities can be found as the solution of the corresponding Schrodinger Equation, along with a *zero potential energy* input to this latter equation:

---

[*] The corresponding author is T. Yarman.

$$E_n = \frac{h^2 n^2}{8 m_0 L^2}, \quad n = 1, 2, 3, \ldots . \tag{1}$$

*(quantum mechanical total energy of the
non-relativistic particle in an infinitely high box)*

Note that this energy amounts to the *kinetic energy* of the particle at the nth energy level, since the potential energy input to the Schrodinger Equation is assumed to be zero. Now we propose to calculate the work done on the particle at the nth energy level in consideration, to compress in it into a smaller volume. This work should lead an increase in the particle's energy:

$$dE_n = -2 dL \frac{h^2 n^2}{8 m_0 L^3} = -2 E_n \frac{dL}{L} . \tag{2}$$

*(increase in the total non-relativistic energy
due to the infinitely small squeezing of the box)*

Since dL is a *negative quantity,* $dE_n$, is as expected, a *positive quantity.*

Let us check whether this energy, turns really out to be the work dW, i.e.
$$dW = p dV = p(-dL) \times (1 \, cm^2) , \tag{3}$$

*(work necessary to furnish to the system
in order to squeeze it, infinitely shortly)*

we would have to furnish to the system to squeeze it as much as –dL; here p is the pressure exerted by the *single* particle on the side walls, as it bounces back.

Recall that the force $f_x$ exerted on the wall perpendicular to the *x* axis, by the particle of mass $m_0$ and velocity $v_x$ (along the x axis, is given as usual, by Newton's second law as

$$f_x = -\frac{\Delta(m_0 v_x)}{\Delta t} , \tag{4}$$

where $\Delta(m_0 v_x)$ is the algebraic increase in the momentum, whilst the molecule bounces back from the given wall, and

$$\Delta t = \frac{2L}{v_x} . \tag{5}$$

Thus $f_x$ becomes
$$f_x = \frac{m_0 v_x^2}{L} . \tag{6}$$

The pressure p exerted by the particle on the wall, is then
$$p = \frac{f_x}{1 \, cm^2} , \tag{7}$$

*(pressure exerted by the particle on the walls)*

since we have assumed that the wall has an area of $1 \, cm^2$.

Via Eq.(3), this makes that the energy we have to furnish to the system to squeeze it as much as –dL, is

$$dW = \frac{m_0 v_x^2}{L}(-dL)(1\,cm^2) \ . \tag{8}$$

*(work we have to furnish to the system
in order to squeeze it, infinitely shortly)*

We specified that the kinetic energy K of the particle in the non-relativistic case, i.e.

$$K = \frac{1}{2} m_0 v_x^2 \ , \tag{9}$$

amounts to the *total energy of the particle* given by Eq.(1).

Eq.(8), then, via Eqs. (9), (1) ad (2) leads to

$$dW = -2E_n \frac{dL}{L} = dE_n \quad (c.q.f.d.). \tag{10}$$

*(work necessary to furnish to the system in order to
squeeze it infinitesimally, amounting to the related
change in the total energy of the particle)*

As trifling as it may seem, to secure the completeness of the presentation, this is still worth to be stated as a theorem.

**Theorem 1:** The work one would furnish to squeeze adiabatically the size of the box in which a non-relativistic, quantum mechanical particle, free of any field, is confined, is fully transformed into an increase in the kinetic energy, or the same, into an increase in the translational energy of the particle.

Many of us could have even assigned the above problem as a homework to our students, although the authors of this paper, do not recall any place, it has been reported to. At any rate, it will constitute the basis of the continuation of our analysis, we present right below.

## 2. THE RELATIVISITIC PARTICLE IN THE BOX: THE WORK DONE TO SQUEEZE THE SIZE OF THE BOX, IS FULLY TRANSFORMED INTO EXTRA RELATIVISTIC KINETIC ENERGY

In case, the particle in the box, is a relativistic particle, the total energy $\Gamma_n$ [cf. Eq.(1)], though, now including the rest mass of the particle can be written as[1]

$$\Gamma_n = mc^2 = \sqrt{m_0^2 c^4 + \frac{c^2 h^2 n^2}{4L^2}} \ , \ n = 1, 2, 3, .... \ ; \tag{11}$$

*(quantum mechanical total energy of the
relativistic particle in an infinitely high box)*

here $mc^2$ is the overall relativistic energy of the particle at the nth state.

Eq.(10) can be expressed as

$$\Gamma_n = \sqrt{m_0^2 c^4 + \varepsilon_n^2} = m_0 c^2 \sqrt{1 + \frac{\varepsilon_n^2}{m_0^2 c^4}} \cong m_0 c^2 + \frac{h^2 n^2}{8 m_0 L^2}, \qquad (12\text{-a})$$

where we have

$$\varepsilon_n = \frac{chn}{2L}, \qquad (12\text{-b})$$

and the last equality of Eq. (12-a), relates to the case where $\varepsilon_n^2/(m_0^2 c^4)$ is negligible as compared to unity; note that $\varepsilon_n$ bears the dimensions of *energy.*

Recall that, for the well in consideration, the de Broglie relationship in regards to the wavelength to be associated with the nth state, shall be written as

$$\lambda_n = 2L = \frac{nh}{mv_x}, \qquad (13)$$

*(nth state wavelength*

so that the quantity $\varepsilon_n$ that we had introduced, can in fact be written as

$$\varepsilon_n = c\mathcal{P}_n, \qquad (14)$$

where $\mathcal{P}_n$ is the momentum $mv_x$ of the particle, at the given state.

When one squeezes the size L, as much as -dL, the total relativistic energy $\Gamma_n$ varies [cf. Eq.(2)] as much as

$$d\Gamma_n = \frac{-\dfrac{dL c^2 h^2 n^2}{4L^3}}{\sqrt{m_0^2 c^4 + \dfrac{c^2 h^2 n^2}{4L^2}}} = -\frac{dL}{L}\frac{\varepsilon_n^2}{mc^2} \cong -2dL\frac{h^2 n^2}{8 m_0 L^3} = -2E_n \frac{dL}{L} = dE_n . \qquad (15)$$

*(increase in the overall relativistic energy
due to the infinitely small squeezing of the box)*

Thus, for a small $\varepsilon_n^2/(m_0^2 c^4)$, as expected, we end up with the result we have derived within the frame of the non-relativistic case. Now, let us calculate the work dW [cf. Eqs. (3), (6) and (8)], to achieve the squeezing process in question:

$$dW = \frac{mv_x^2}{L}(-dL)(1\text{cm}^2) ; \qquad (16)$$

*(work necessary to furnish to the system in order to
squeeze it infinitely shortly, amounting to the related
change in the overall relativistic energy of the particle)*

where though we had to use the relativistic mass m, instead of the rest mass we have used in Eq.(3).

Via Eqs. (12-ba), (13), (14) and (15), this can be written as

$$dW = -\frac{m^2 v_x^2}{mL} dL = -\mathcal{P}_n^2 \frac{dL}{mL} = -\frac{dL}{L}\frac{\varepsilon_n^2}{mc^2} = d\Gamma_n . \qquad (17)$$

Thus, in the case of a relativistic particle too, we well check out the result, derived previously, based on the non-relativistic particle. Thus we can state the following theorem.

**Theorem 2:** The work one would furnish to squeeze adiabatically the size of an isolated box in which a relativistic, quantum mechanical particle, free of any field, is confined, at a given energy level, is fully transformed into a change in the relativistic energy of the particle, at the given energy level.

Now, the *relativistic kinetic energy* $K_n$ of the particle at the nth level, as usual, is

$$K_n = mc^2 - m_0 c^2 = \Gamma_n - m_0 c^2 . \qquad (18)$$
*(relativistic kinetic energy of the particle)*

Thus, via differentiating, and Eq.(17),

$$dK_n = c^2 dm = d\Gamma_n = dW . \qquad (19)$$

This means that, all the work furnished to the particle, through the squeezing process, is transformed into *extra kinetic energy* the particle would acquire. It is worth to specify this result as a next theorem.

**Theorem 3:** The work furnished to squeeze adiabatically the size of an isolated box in which a relativistic, quantum mechanical particle, free of any field, is confined at a given energy level, is fully transformed into extra kinetic energy, the particle will acquire, which thus conjointly increases the relativistic mass of the particle just as much, at the given energy level.

We show elsewhere that, an *"ideal gas"* behaves in accordance with the foregoing derivation, were the gas molecules can indeed be considered independently from each other.[2]

Let us summarize the approach in question.

## 4. THE WELL-MATCH OF THE LAW OF GASES WITH QUANTUM MECHANICS BASED ON THE CONSTANCY OF $PV^\gamma$ FOR AN *ADIABATIC TRANSFORMATION*

The relationship
$$PV^\gamma = \text{Constant} , \qquad (20)$$

for an adiabatic transformation, an *"ideal gas"* undergoes, constitutes an interesting check point of the *compatibility* of the macroscopic laws of gases, and quantum mechanics.[8] Here P is, as usual, the pressure the gas of volume V exerts of the walls of the container in consideration. Below, for simplicity, we will operate with one mole of gas.

Thus, the first author et al., have previously proposed to calculate specifically, the constant in question, within a quantum mechanical framework.[2]
.
Eq.(20) involves, the usual definition[3]

$$\gamma = \frac{C_P}{C_V}, \tag{21}$$

where

$$C_V = \frac{3}{2}R, \tag{22}$$

$$C_P = \frac{5}{2}R; \tag{23}$$

$C_V$ is the heat to be delivered to one mole of ideal gas at constant volume to increase its temperature as much as 1° K, and $C_P$ being the heat to be delivered to one mole of ideal gas at constant pressure to increase its temperature as much as 1° K; R is the gas constant. Eqs. (22) and (23) are *exact,* when internal energy levels of molecules are not excited. This approximation, is classically, fulfilled for an *ideal gas,* by definition. Thus we have

$$\gamma = \frac{5}{3}. \tag{24}$$

Since in an *ideal gas,* all molecules are assumed to behave independently from each other, Eq.(21) remains still valid, had the gas consisted, even in one single molecule. It may indeed be recalled that, within the frame of the *kinetic theory of gases,* one first expresses the pressure for just one molecule of gas, before he proceeds for very many more, making up the gas of concern. This is how, one formulates the *macroscopic pressure,* the gas exerts on the walls of its container [cf. Eqs. (4), (5), (6) and (7)].

**The Non-Relativistic Case**

Let us consider a particle of rest mass $m_0$ with a fixed internal energy state, located in a macroscopic cube of side L. The non-relativistic Schrödinger equation furnishes the nth energy $E_n$ [cf. Eq. (1)]:

$$E_n = \frac{h^2}{8m_0}\left(\frac{n_x^2}{L^2} + \frac{n_y^2}{L^2} + \frac{n_z^2}{L^2}\right) = \frac{h^2\left(n_x^2 + n_y^2 + n_z^2\right)}{8m_0 L^2}, \tag{25}$$

*(non-relativistic energy in three dimensions)*

where we denoted $n_x = 1, 2, 3, ..., n_y = 1, 2, 3, ..., n_z = 1, 2, 3, ...$, the quantum numbers to be associated with the corresponding wave function dependencies on the respective directions x, y and z. Here, for brevity, while writing $E_n$, we introduced, the subscript "n", to denote the given state characterized by the integer numbers $n_x$, $n_y$, and $n_z$; so each "n" in fact, represents a set of three integer numbers.

For an ideal gas confined, in an infinitely high box, the potential energy input to the Schrodinger Equation, is null, in the inside. Thus, for the non-relativistic case which is generally the case, we have

$$E_n = \frac{1}{2}m_0 v_n^2, \tag{26}$$

$v_n$ being the velocity of the particle at the nth energy level.

At the given energy level, the *pressure* $p_n$ exerted by the *single particle* in consideration, this time in three dimensions, on the walls, becomes [cf. Eqs. (6) and (7)]

$$p_n = \frac{m_0 v_n^2}{3L^3} = \frac{2}{3}\frac{E_n}{L^3}. \tag{27}$$

Now, let us calculate the product $p_n V^\gamma$:

$$p_n V^\gamma = \frac{2}{3}\frac{\dfrac{h^2(n_x^2 + n_y^2 + n_z^2)}{8m_0 L^2}}{L^3}(L^3)^{5/3} = \frac{h^2(n_x^2 + n_y^2 + n_z^2)}{12 m_0}. \tag{28}$$

*(quantum mechanical constancy to be associated with the adiabatic transformation)*

This quantity indeed, turns out to be a constant for a given particle of mass $m_0$ at the given energy level n (specified by the set $n_x$, $n_y$, and $n_z$).

Recall that the total quantized energy $E_n$ of Eq. (25) ultimately determines the quantized velocity $v_n$ of Eq. (26), along with its three components, which are all, in return, quantized.

When it is question of many particles instead of just one, *normally,* we will have particles at different, possible, quantized states. This is *most likely* what leads to the Maxwellian distribution of particles, with different translational energies, in a given container, at a *given temperature.*

We can anyway visualize the *average particle,* at the $\overline{n}$ th level, thus corresponding to the *temperature* coming into play, and well suppose that all others, behave the same. Furthermore, all three components of the *average velocity, in equilibrium* will be expected to be the same. Thus, we can rewrite Eq.(28) for the macroscopic pressure $P_{\overline{n}}$ exerted at the given *average state* $\overline{n}$, by one mole of gas, on the walls of the container:[†]

---

[†] Rigorously speaking,[2] one must write

$$P_{\overline{n}} V^\gamma = \sum_{i=1}^{N_A} \frac{h^2(n_{ix}^2 + n_{iy}^2 + n_{iz}^2)}{12m} = N_A \frac{h^2 \overline{n^2}}{4m}, \tag{i}$$

along with the definition,

$$\frac{1}{3N_A}\sum_{i=1}^{N_A}(n_{ix}^2 + n_{iy}^2 + n_{iz}^2) = \overline{n^2}. \tag{ii}$$

Thus it becomes clear that, if all particles beared the same set of quantum numbers, each with equal quantum numbers along all three directions, i.e. $n_x = n_y = n_z = \overline{n}$, then $\overline{n}$ becomes

$$\overline{n} = \sqrt{\overline{n^2}}. \tag{iii}$$

$$P_{\bar{n}} V^{\gamma} = N_A \frac{h^2 \bar{n}^2}{4m_0} ; \qquad (29)$$

*(the adiabatic constancy)*

$N_A$ is the Avagadro Number. Thus, Eq.(29) discloses the constant involved by the *adiabatic transformation relationship,* i.e. Eq.(20). Note that, at the average state $\bar{n}$ *(i.e. at the given temperature),* the *mean square speed* of the gas molecules is $v_{\bar{n}}^2 = \overline{v_n^2}$ ; the average energy $E_{\bar{n}} = \overline{E_n}$ is furnished accordingly, via the framework of Eq.(26).

The above finding is worth to be stated as our next theorem.

**Theorem 4:** The constancy $P_{\bar{n}} V^{\gamma}$, drawn by an adiabatic transformation of an ideal gas, is nothing but a *macroscopic quantum mechanical behavior.* The constant, for one mole of gas becomes $N_A h^2 \bar{n}^2 /(4m_0)$, where $N_A$ is the Avagadro Number, h the Planck Constant, $\bar{n}$ the quantum number characterizing the corresponding average state, and $m_0$ the mass of molecules making up the gas.

Let us now undertake the same problem, though highly improbable, in the *relativistic case.*

**The Relativistic Case**

The pressure $p_n$ delineated by just one molecule, at a given nth level, is given by Eq.(27), where how ever now, we have to consider the relativistic mass m, furnished by Eq.(10):

$$p_n = \frac{mv_n^2}{3L^3} = \frac{m^2 v_n^2}{3mL^3} = \frac{\mathcal{P}_n^2}{3mL^3} , \qquad (30)$$

here $\mathcal{P}_n$ is the *relativistic momentum* furnished by Eqs. (12) and (14), so that

$$p_{\bar{n}} = \frac{\mathcal{P}_{\bar{n}}^2}{3mL^3} = \frac{h^2 \left(\overline{n_x^2} + \overline{n_y^2} + \overline{n_z^2}\right)}{(3mL^3)(4L^2)} = \frac{h^2 \bar{n}^2}{4mL^5} , \qquad (31)$$

where the last equality, is written in the case we consider the three quantum numbers to be equal to each other, for the average state we characterize by $\bar{n}$, just the same way we did, through the previous *non-relativistic approach.*

We can then compose $P_{\bar{n}} V^{\gamma}$ for one mole gas [cf. Eq.(30)]

$$P_{\bar{n}} V^{\gamma} = N_A \frac{h^2 \bar{n}^2}{4mL^5} \left(L^3\right)^{5/3} , \qquad (32)$$

which, via Eq.(10), leads to

$$P_{\bar{n}} V^{\gamma} m_0 \sqrt{1 + \frac{h^2 \bar{n}^2}{4 m_0^2 c^2 L^2}} = N_A \frac{h^2 \bar{n}^2}{4} . \qquad (33)$$

*(quantum mechanical constancy to be associated with the
adiabatic transformation of a gas made up high speed molecules*

Thence, we come to the conclusion that in the relativistic case, $P_{\bar{n}} V^{\gamma}$ does not remain constant, since the multiplier of this term, displayed at the LHS of the above relationship, depends on the size of the container of concern *(albeit the RHS well appears to be a constant)*.

At any rate, as clearly seen [cf. Eqs. (11) and (33)], the quantity

$$mc^2 P_{\bar{n}} V^{\gamma} = \Gamma_{\bar{n}} P_{\bar{n}} V^{\gamma} , \qquad (34)$$

for an adiabatic transformation, remains constant. We state this interesting occurrence, as our next theorem.

**Theorem 5:** In a gas where molecules would bare the relativistic energy $\Gamma_{\bar{n}} = mc^2$, in the average, at the $\bar{n}$ th state, the product $\Gamma_{\bar{n}} P_{\bar{n}} V^{\gamma}$, through an adiabatic transformation, remains constant, which, for one mole of gas, becomes $N_A h^2 \bar{n}^2 / 4$; $N_A$ is the Avagadro Number, and h the Planck Constant. The adiabatic constancy of the quantity $\Gamma_{\bar{n}} P_{\bar{n}} V^{\gamma}$ is nothing but a macroscopic quantum mechanical manifestation, the gas delineates.

## 4. GENERALIZATION: AN IDEAL GAS IS MADE OF INDIVIDUAL QUANTUM CONSTITUENTS, IMPRISONED IN THE CONTAINER OF THE GAS, WELL YIELDING TO THE KNOWN MACROSCOPIC THERMODYNAMIC MANIFESTATIONS

Theorems 4 and 5, can be extended as follows.

**Theorem 6:** A gas, in which molecules do not interact with, is structured in accordance with the *quantum mechanical behavior* of just a single particle imprisoned in its container.

This, right away, induces the following two essential theorems.

**Theorem 7:** The well established thermodynamic behavior of an ideal gas is nothing but an overall macroscopic manifestation of the quantum mechanical conduct of each of its constituents, imprisoned in the container in consideration.

**Theorem 8:** An ideal gas, as defined classically, is in fact a gas made of constituents, each behaving individually, and as a quantum mechanical particle imprisoned in the container of the gas of concern.

This latter theorem thus provides a quantum mechanical foundation to the *classical definition of an ideal gas,* classically based on the empiric law of gases, i.e.

$$PV=RT, \qquad (35)$$

written for just one mole of gas, where as usual, P is the pressure the gas exerts on the walls of its container of volume V, at the absolute temperature T, and R is the gas constant (i.e. 8.31 Joules / ° K).

Hence, we can generalize Our Theorems 6, 7 and 8, proven for a single quantum mechanical object, imprisoned in a box, to the case of an *ideal gas,* thus made of non-interacting quantum mechanical constituents.

**Theorem 9:** The work one would furnish to squeeze adiabatically an ideal gas, thus made of non-interacting quantum mechanical particles, assumed to be free of any field, amounts to the related change in the *average kinetic energy* of the particles, concomitantly increasing the *relativistic mass* of the constituents. Conversely the adiabatic expansion of a gas, delivers to the outside, the energy stored, as extra translational energy, within the box.

Once said, this theorem may seem to be trivial. Nonetheless we have to recall that, we were able to state it, generally, i.e. regardless how fast the constituents making the gas, may move.

It was thus important to elucidate the fact that, the gas constituents all behave individually, as quantum mechanical objects, imprisoned in the container of the gas. *"Individuality"* evidently, excludes any possible interactions in between the gas constituents. This point is not only crucial in avoiding *possible non-linearities,* but is also crucial in dropping all kinds of electric field interaction between the gas constituents, despite the ordinary repulsion that can take place in between the electronic clouds of these. At any rate, the work $\Delta W$ furnished adiabatically to an ideal gas, can indeed be translated as extra relativistic energy

$$\Delta W = c^2 \Delta m, \qquad (36)$$

where $\Delta m$ represents the average relativistic mass increase of the constituents, to be stored as additional kinetic energy, or the same, translational energy of these [cf. Eq.(10)]

Based on the foregoing discussion, below, we would like to handle a fundamental problem.

## 5. IF WORK IS ACHIEVED BETWEEN THE TWO COMPARTMENTS OF A *CLOSED SYSTEM,* WHAT HAPPENS TO THE OVERALL INTERNAL ENERGY?

Suppose we have two adjacent compartments filled disjointedly with an ideal gas, under different pressures $P^{Higher}$ and $P^{Lower}$, respectively, and separated by an impermeable, how ever *mobile* wall, when set free. For simplicity, let us assume that both compartments contain the same number of gas molecules *(thus, at different temperatures).* The whole apparatus constitutes a totally isolated system. Let V the volume of the compartment, where the high pressure gas is imprisoned.

If the wall in the middle, unlocked, the gas with *higher pressure* will achieve *work*, pushing away the separation wall, as much as dV *(which is, manifestly, a positive quantity)*. The gas with *higher pressure* will then expand, and loose energy. Conjointly, the gas with *lower pressure* $P^{Lower}$ will receive the work coming into play. Thus, it will get compressed, and receive energy. Let us suppose that there is no energy loss at the wall mechanism, in the middle of the two compartments of gas. The overall work $\delta W$ coming into play is, as usual,

$$\delta W = \left(P^{Higher} - P^{Lower}\right) dV . \tag{37}$$

*(work achieved by the gas of higher pressure,*
*expanding as much as* dV, *on the gas of lower pressure)*

Clearly, a *net positive amount of work* is displayed. Since the whole system is *isolated,* this work is not furnished to the outside. The system, being totally isolated, has not either been exposed to any outside heat source. Hence, the *variation of its internal energy,* classically speaking, must be *zero*. Yet, there indeed came into play, a *"net positive work"* $\delta W$, which seemingly should have led to a *variation in the internal energy,* accounting for it. Thereby, at *a first strike,* we seem to end up with a paradox.[4] How do we handle such as problem, based on the discussion presented above?

To make our analysis, as clear as possible, let us reason on just *one constituent* tagged at each compartment *(given that the gas, macroscopic behavior, as we have shown, is nothing but a simple superposition of the independent behavior of all of the constituents making it, all supposed to be at the same average energy level)*. Then, the answer of the above question, after all, turns out to be simple. Theorem 9, thus fundamentally, Theorems 6, 7 and 8, tell us that the *kinetic energy* $E_{n0}^{Higher}$ of a constituent *originally* at the nth level, in the *high pressure* compartment, representing the average object of the expanding gas, will decrease as much as $\delta w$, work achieved by the single object in question. Thus, through the expansion process in question, $E_{n0}^{Higher}$ becomes $E_{n}^{Higher}$, so that

$$E_{n}^{Higher} = E_{n0}^{Higher} - \delta w . \tag{38}$$

*(kinetic energy of an average individual constituent at the nth level*
*in the high pressure gas, following an infinitesimal expansion)*

Recall that $\delta w$ is a *positive quantity*. Recall further that, classically speaking $E_{n0}^{Higher}$, is the internal energy $U_{n0}^{Higher}$ of the gas of higher pressure, *per constituent.*

That is

$$U_{n0}^{Higher} = E_{n0}^{Higher} . \tag{39}$$

*(internal energy of the average constituent at the*
*nth level, in the high pressure compartment)*

As mentioned, supposing there are *equal numbers* of gas constituents at both compartments *(as we have adopted, to simplify our reasoning),* the kinetic energy $E_{m0}^{Lower}$ of the constituent originally at the mth level, in the *low pressure* compartment, representing the *average object* of the compressed gas, will increase as much as $\delta w$ *(the work produced by the object in question, in the high pressure compartment)*. Thus, $E_{m0}^{Lower}$ becomes $E_{m}^{Lower}$, so that

$$E_m^{Lower} = E_{m0}^{Lower} + \delta w . \tag{40}$$
*(kinetic energy of an average individual constituent in the
low pressure gas, following the infinitesimal expansion)*

Recall that, classically speaking, $E_{m0}^{Lower}$ is the internal energy $U_{n0}^{Lower}$ of the gas of low pressure, *per constituent*. That is

$$U_{m0}^{Lower} = E_{m0}^{Lower} . \tag{41}$$
*(internal energy of an average constituent at the
mth level in the high pressure compartment)*

Under these circumstances, the *overall internal energy* $U_{n,m}$ of the closed system, *per constituent* is then

$$U_{n,m} = U_m^{Lower} + U_n^{Higher} . \tag{42}$$

Let us now write down, how the quantities appearing in this equation, change. Thus, the *algebraic change in the internal energy* $dU_n^{Higher}$, for the *average constituent*, at the nth energy level, at the high pressure compartment, is

$$dU_n^{Higher} = E_n^{Higher} - E_{n0}^{Higher} = -\delta w . \tag{43}$$
*(algebraic change in the kinetic energy of an
individual constituent in the high pressure
compartment following an infinitesimal expansion)*

Similarly, the *algebraic change in the internal energy* $dU_m^{Lower}$, for the *average constituent*, at the mth energy level, at the low pressure compartment, can be written as

$$dU_m^{Lower} = E_m^{Lower} - E_{m0}^{Lower} = \delta w . \tag{44}$$
*(algebraic change in the kinetic energy of an
individual constituent in the high pressure
compartment following an infinitesimal expansion)*

Thus, the *overall change in the internal energy* $dU_{n,m}$, the *average constituent*, at respectively n and m levels, at the two compartments, delineate, can be written as [cf. Eqs. (42), (43) and (44)]

$$dU_{n,m} = dU_m^{Lower} + dU_n^{Higher} = \delta w - \delta w = 0 . \tag{45}$$
*(overall internal energy change of the system, per constituent)*

Thence the variation of the overall internal energy remains identically zero, meaning that the overall internal energy remains constant. Therefore, Eq.(45), overall, is nothing but, the expression of the *law energy conservation law*.

In conclusion, for the ideal gas case *(made of non-interacting, quantum mechanical objects)*, under consideration, there is no *paradox*, on he contrary to what was formulated right above.

That is, in the isolated system, although there happens a finite work exchanged between the two compartments, the *overall change in the internal energies* per constituent, of the gases, *i.e. the sum of* $dU_n^{Higher}$ and $dU_m^{Lower}$, well amounts to zero, as expected. The original internal energy per constituent, $U_{n0}^{Higher}$ of the gas with high pressure decreases. And the original internal energy per constituent, $U_{n0}^{Lower}$ of the gas with lower pressure increases. But the overall internal energy remains the same.

It would be interesting to revisit Joule's experiment,[5] where Joule demonstrated that the internal energy of a gas, depends only on the temperature, and not for instance on the volume of the gas, from the angle we have drawn herein. But because peculiarities to be elaborated within the frame of this experiment, we save that study for a further work.

Let us emphasize that, throughout the foregoing discussion, we have overlooked any possible interaction between the constituents making up the gas; we have indeed supposed that these constituents all behave independently from each other. At a first glance it does not anyway seem easy to handle the case where one would allow an interaction between molecules making up a gas. Too much pressure put on a gas, would necessarily bring in, *electrostatic repulsion effects,* which would involve *energy stored as an infinitesimal increase of rest masses.* The expansion of the gas, conversely would involve the transformation of *extra rest mass into kinetic energy.*

**CONCLUSION**

Herein we come to disclose that, the *ideal gas behavior* is in fact nothing else, but a *quantum mechanical manifestation, at the macro level.* This allowed us to define the *ideal gas,* based on the behavior of constituents making it up, thus each behaving as a quantum mechanical particle in the box, not interacting with each other, whatsoever. The *box* at the macro level is the *container* imprisoning the gas in consideration. Thus, the *adiabatic* compression of a gas yields the increase of the *quantum mechanical relativistic kinetic energy* of its constituents, moving in a zero field, inside the container. This disclosure allowed us, to answer the following question, seemingly paradoxical, at a first strike: *If work is achieved between the two compartments of a closed system, would the overall internal energy, change?* The answer is as follows. The internal energy of the gas confined in the compartment of higher pressure decreases. The internal energy of the gas confined in the compartment of lower pressure increases. But the overall internal energy is kept constant. This is in fact nothing else, but the plain law of conservation of energy.

Too much pressure put on a gas, would necessarily bring in electrostatic repulsion effects, which would involve energy stored as *mass equivalent.* The expansion of the gas, conversely would involve the transformation of mass into kinetic energy. We save such an approach for a subsequent article.

**REFERENCES**

1. J. C. Davis, Jr., Chapter 6, Advanced Physical Chemistry, The Ronald Press Company, New York, 1965.

2. T. Yarman, A. L. Kholmetskii, Ö. Korfali, On A Link Between Classical Phenomenologıcal Laws of Gases, And Quantum Mechanics, arXiv:0805.4494, 2008, http://arxiv.org/ftp/arxiv/papers/0805/0805.4494.pdf.

3. A. Sommerfeld, Thermodynamics and Statistical Mechanics, Academic Press, New York, 1964.

4. V. Krasnoholovets, J-L Tane, An extended Interpretation of the Thermodynamic Theory, Including an Additional Energy Associated with a Decrease in Mass, International Journal of Simulation and Process Modelling, Vol. 2, Nos. 1/2, 67, 2006.

5. J. P. Joule, Phil. Mag., 1843, 23, p. 263.